
\def\tfr{{Tully-Fisher relation}}

\def\sect{\vskip 5mm \centerline}
\def\section{\vskip 5mm \centerline}

\def\r{\hangindent=1pc  \noindent}
\def\ref{\hangindent=1pc  \noindent}
\def\cen{\centerline}

\def\v{\vskip 2mm}

\def\endpage{\vfil\break}

\def\v{\vskip 2mm}

\def\no{\noindent}
\def\kms{km s$^{-1}$}
\def\deg{$^\circ$}

\def\Tmb{T_{\rm mb}}

\def\kms{km s$^{-1}$}
\def\Msun{M_{\odot \hskip-5.2pt \bullet}}

\def\lsun{$L_{\odot \hskip-5.2pt \bullet}$}

\def\deg{$^\circ$}

\def\co{$^{12}$CO($J=1-0$)}

\def\aa{{\it Astron. Astrophys.}}

\baselineskip=16pt

\no
{\bf  CO versus HI in the Tully-Fisher relation for a sample of
32 galaxies}\v

\no
Franz Sch\"oniger (1)  \& Yoshiaki Sofue (1)

\no
{(1) Institute of Astronomoy, University of Tokyo, Mitaka, Tokyo 181}

\v
\v

\no
F. Sch\"oniger \& Y. Sofue:
CO versus HI in the Tully-Fisher relation

\v
\no
{\it Send offprint requests to:} Y. Sofue

\vskip 3cm
\no
Main Journal

\no
Extragalactic astronomy

\no
07.09.1 Galaxies: general

\no
04.03.1 Distance: distance scale

\no
18.05.1 Radio lines: 21 cm

\no
18.07.1 Radio lines: molecular

\no
Proofs to Y. Sofue

\vfill
\eject

\no

{\bf Abstract.}
As a basic step for establishing the CO-line
\tfr\ for distant galaxies, we
made a comparative study of HI versus CO line profiles.
Total line profiles of the CO line emission from 32 galaxies
have been compared with the corresponding HI emission.
We found a good correlation between
the profiles of CO and HI. This strongly
supports the thesis that CO can be used
in the Tully-Fisher relation for distant galaxies
where HI observations cannot reach.
We also argue that CO can be used as an alternative
to HI for un-isolated galaxies such as those in dense
cluster of galaxies.
Using the B-magnitudes
and a recent calibration of the HI \tfr
we give the distances for the galaxies
derived from the CO-line Tully-Fisher relation and compare them with the
corresponding HI distances.

\v
\no
{\bf Keywords:}  Galaxies: general\ -\ Distances: distance scale
  \ -\ Radio lines: 21 cm  \ -\ Radio lines: molecular

\v

\no
{\bf 1. Introduction}\v

The \tfr\ is one of the most powerful tools to estimate distances
to galaxies (Tully \& Fisher 1977; Aaronson et al. 1986;  Pierce
\& Tully 1988;
Kraan-Korteweg et al. 1988;  Fouqu{\'e} et al. 1990; Fukugita et al. 1991).
Inspite of many different
photometric techniques (see the literature above), almost identical HI data
have been used (e.g., Bottinelli et al. 1989;  Huchtmeier et al. 1989).
Distances to galaxies so far reached by HI observations are limited to
around 100 Mpc, or recession velocities of 10,000 to 15,000 \kms\ even
with the use of the largest telescope (see, e.g., Dickey \& Kazes 1992).
However, we have no routined method to detemine distances to galaxies beyond
this distance, except for the possible use of CO line using
a large-aperture mm-wave telescopes.

Beyond this distance, angular resolutions of a few arcminutes
in HI observations becomes too large to resolve individual
galaxies in a cluster.
Interferometers like VLA are not useful for the purpose
because of the limited number of spectral channels (velocity resolution)
due to the limited number of auto-correlator channels.
Furthermore, more red-shifted HI frequency results in
increases in beam size as well as in interferences,
which also makes resolutin of distant cluster galaxies difficult.

On the other hand, CO facilities have much sharper beams
(e.g., 15 arcsec with the Nobeyama 45-m telescope), with the
use of which we would be able to resolve individual member galaxies
in a cluster more easily, making it possible to avoid
contamination by other member galaxies in a beam.
Moreover, the larger is the redshift of an object, the lower becomes the
CO frequency, which results in a decrease in the system noise temperature
due to atmospheric O$_2$ emission near 115 GHz:
the more distant is a galaxy, the lower becomes the noise temperature.
Only the disadvantage of the use of CO line would be its sensitivity,
particularly for distant galaxies.
Actually, we need a few mK rms data for line-width measurements
with a velocity resollution of 10 \kms\
for normal galaxies beyond $cz \sim 10,000$ \kms,
for which an integration time of about ten or more hours are required,
and such observations are possible only by a long-term project
with the largest mm-telescopes (Sofue, Sch{\"o}niger, Kazes, and Dickey,
private communication).
However, although noise temperatures of current CO receivers,
which are some hundred K in the present status,
are still worse than those used in HI observations,
we have the hope that they will be much improved in the near future
to several tens K, which may result in an increase of the sensitivity
by one or two orders of magnitudes.

Since the distribution of HI gas is broad in a galaxy,
HI line profiles are easily disturbed by interactions among galaxies,
which is inevitable in the central region of a cluster.
Such disturbance might cause uncertainty in
the HI line profiles for \tfr.
On the other hand, CO gases are more tightly correlated with
the stellar distribution, and are less affected by the
tidal interaction.
Of course, CO gas is distributed enough to a radius
of several to ten kpc, so that the integrated line profiles manifest the
maximum velocity part of the rotation curve (Sofue 1992).

Recently, Dickey and Kazes (1992) addressed the question whether
linewidths in the CO line emission ($\lambda=$2.6 and 1.3 mm) can
be used in the Tully-Fisher relation as  an alternative or
supplement to HI.
They found a linear correlation between CO and HI linewidths
for the Coma and other nearby
clusters of galaxies  and showed the hope that
the CO Tully-Fisher relation
can be useful for distant clusters of galaxies.
We also obtained a good correlation between
CO and HI for our sample of four isolated and interacting
edge-on galaxies (Sofue 1992).
We suggested that the CO-line
can be used not only for \tfr\ on normal galaxies, but also for
interacting galaxies, which is particularly important when
galaxies sitting in the central region of a cluter are concerned.

In this paper now we show a detailed comparison of CO and HI
profiles for
32 spiral and irregular field  galaxies.
This is the basic step toward establishing the CO \tfr\ not only for
distant galaxies using the Nobeyama 45-m mm-wave telescope,
but also for nearby glaxies as well as
for galaxies in the central region of clusters where
tidal interaction is inevitable.

Many of our sample galaxies will
show peculiar features like active nuclei
or interaction with a companion.
We focus on the question whether, despite such peculiarities,
there is still a correlation between CO and HI. Further we
investigate the question how strongly
the eventually occurring differences between the profiles
influence the resulting distances from the \tfr.

Then we apply the  calibration of the B-band
Tully-Fisher relation  to determine the
distances of the galaxies using the CO linewidth. We thereby
use the CO linewidths just instead of HI without any own calibration.
Finally we compare the resulting distances with the distances obtained
from the HI \tfr.
\v
\no
{\bf 2. Data}\v

Our sample of galaxies is listed in Table 1. Column one gives the galaxy
name, column two the inclination i and column three the total
"face-on" blue magnitude $B_{0}^{\rm T}$
corrected for galactic and internal absorption
which was taken from de Vaucouleurs et al. (1991).
\v
\no
Table 1.
\v

With the exception of NGC 891, NGC 1808, NGC 3079,
 NGC 4565, NGC 4631 and NGC 5907 which have been
 observed in \co using the
Nobeyama 45 m telescope,  all the  line
profiles of CO and HI are taken from the literature. If the
galaxy has been mapped in CO we have added up the individual
spectra to get the total line profile.
Figure 1  shows
 the total line profiles of CO (solid lines)
 for our sample with the
corresponding HI profiles (dashed lines) superimposed.
We measured the linewidths at the 20\% and 50\% level.
In the following we give some notes on the individual galaxies.

NGC 224 (M 31):
This is a nearby Sb galaxy (D=0.78 Mpc) and
one of the local calibrators for
the \tfr. It has been mapped by Koper et al. (1991).
In their
paper they also gave a comparison between the HI and CO profiles
which shows an almost perfect agreement between CO and HI.
The good agreement between the linewidths (${W_{20}^{\rm CO}=535}\pm 10$ km/s,
${W_{20}^{\rm HI}=540}\pm 10$ km/s)
would lead to only -1 \% difference in distances $\Delta \rm D$, whereas
$\Delta \rm D$ is defined as $({D_{\rm CO}-D_{\rm HI})/D_{\rm HI}}$.

NGC 520, NGC 660, NGC 2623 and NGC 7674:
The CO profiles of these galaxies are  taken from  Sanders \&
Mirabel (1985), the HI profiles are from Mirabel \& Sanders (1988).
They are all suspected of being strongly
interacting contact pairs.
NGC 520 (${W_{20}^{\rm CO}=360}\pm 20$ km/s,
${W_{20}^{\rm HI}=420}\pm 10$km/s), shows a narrower CO profile
at 20\% level, but broader when compared at 50\% level.
NGC 660 (${W_{20}^{\rm CO}=370}\pm 20$ km/s,
${W_{20}^{\rm HI}=330}\pm 10$ km/s) and
 NGC 2623 (${W_{20}^{\rm CO}=450}\pm 20$ km/s,
 ${W_{20}^{\rm HI}=400}\pm 10$ km/s)
show a
broader profile in CO. In the case of
NGC 7674
(${W_{20}^{\rm CO}=190}\pm$20 km/s, ${W_{20}^{\rm HI}=215}\pm$20 km/s)
the HI profile in emission looks narrower, but total width
including the absorption feature is broader than CO.
The corresponding diffferences in distances for $\Delta D$ for these
galaxies
are -18\% for NGC 660, 17\% for NGC 2623,
-18\% for NGC 7674 and -23\% for NGC 520.

NGC 891:
This is a typical and ``standard'' edge-on galaxy
 of Sb type, and is isolated and not disturbed (e.g. Sandage 1961).
We find an almost perfect coincidence between the CO and
 HI profiles, except for a slight asymmetry with respect to
the systemic velocity: both the lines show an almost identical
double horn profile.
The linewidths at 20\% of peak intensity
(${W_{20}^{\rm CO}=W_{20}^{\rm HI}=490}\pm10$km/s)
are  the same within the errors. The HI profile was taken from Rots (1980).

NGC 992 and NGC 7469:
These galaxies also have been observed in CO by Sanders \& Mirabel (1985).
the HI profiles
are taken from Mirabel \& Sanders (1988).
For NGC 992 the CO profile looks rather peculiar but the linewidths
between CO and HI (${W_{20}^{\rm CO}=385}\pm 20$ km/s,
${W_{20}^{\rm HI}= 360}\pm  10$ km/s) agree quite well.
For NGC 7469 the CO profile is blue-shifted by about 80 km/s
but nevertheless the linewidths (${W_{20}^{\rm CO}=330}\pm 20$ km/s,
${W_{20}^{\rm HI}=370}\pm 10$ km/s) agree approximately.
This leads to a $\Delta  D$ of 11\% for
NGC 992 and -16\% for NGC 7469 respectively.
Hence that NGC 992 as well as NGC 7469 both have a close companion
(Condon \& Condon 1982).

NGC 1068 and NGC 1808:
These  galaxies both have a Seyfert nucleus, in the case of
NGC 1808  additionally a jet is emerging
from the central region (V\'eron-Cetty
\& V\'eron 1985). NGC 1068 has been observed in CO by
Scoville et al. (1983), the HI profile was taken from
Staveley-Smith (1987). NGC 1808 has been observed using
the Nobeyama 45 m telescope by Sofue et al. (1992),
the HI profile was taken from Rots (1980).

Both galaxies show a broader CO profile at the 50\% level
but the linewidths agree pretty well at the 20\% level. This
suggests a falling rotation curve beyond the maximum rotational
velocity for both galaxies.
For NGC 1068 the linewidths (${W_{20}^{\rm CO}=350}\pm 20$ km/s,
${W_{20}^{\rm HI}=345}\pm 10$ km/s) leads to a difference
of 2\% in distances, for NGC 1808 (${W_{20}^{\rm CO}=340}\pm 20$ km/s,
${W_{20}^{\rm HI}=330}\pm 10$ km/s) to a difference of
5\%.

NGC 1365:
This barred spiral galaxy of Sb type has been observed in CO by
Sandqvist et al. (1988). The CO is distributed along the bar
and it's profile (${W_{20}=370}\pm 10$ km/s
is slightly narrower than that
of HI (${W_{20}=390}\pm10$ km/s. The different linewidths lead to
a $\Delta D$ of -9\%.

NGC 1569:
This is an irregular galaxy showing an extremely narrow profile.
A comparison of CO and HI profiles given by Young et al. (1984)
shows an almost perfect agreement between the both species.
The CO linewidth is ${W_{\rm CO}=85}\pm15$ km/s and the corresponding
HI linewidth is ${W_{20}=95}\pm10$ km/s agree within the errors.
Since the linewidths are so narrow even the small difference
in linewidths
leads to a $\Delta D = -25\%$.

NGC 2146:
An Sb galaxy with a narrower profile in CO
(${W_{20}=390\pm}20$ km/s) than in HI (${W_{20}=460}\pm 10$km/s).
The CO profile is taken from Young et al. (1986), the HI profile is from
Tift \& Cocke (1988). The corresponding value of $\Delta  D$ is
-23\%.

NGC 2276:
An infrared bright SBc galaxy which has been observed in CO by
Young et al. (1986). The CO profile (${W_{20}=140}\pm20$ km/s)
is narrower than the corresponding HI profile (${W_{20}=200}\pm20$ km/s)
taken from  Shostak (1978). This corresponds to a $\Delta D$ of
-36 \%.

NGC 2339:
Another infrared bright spiral also taken from Young et al. (1986).
The CO profile (${W_{20}=330}\pm 20$km/s) and the HI profile
(${W_{20}=340}\pm10$ km/s) coincide within the errors,
despite the CO profile is blueshifted by about 70 km/s against HI.
For this galaxy $\Delta  D$ is -5\%.

NGC 3034 (M82):
This famous irregular galaxy has been observed in CO
(${W_{20}=320}\pm20$km/s) by
Young and Scoville (1984), the HI profile
(${W_{20}=300}\pm20$ km/s) is from Crutcher et al. (1978).
Despite the peculiarity of this galaxy the profiles agree approximately.
 The distances derived from CO and HI respectively differ
by $\Delta D = 11\%$.

NGC 3079:
This amorphous edge-on galaxy is known for its vertical
 radio lobes and a variety of activity (e.g., Duric et al. 1983).
The CO gas is highly concentrated near the
 nucleus, displaying a high velocity dispersion
 (Young et al. 1988; Sofue \& Irwin 1992).
We also show a total profile constructed from a
 position-velocity diagram along the major axis
 obtained by the Nobeyama mm Array (NMA) (Sofue and Irwin 1992),
 where the effective beam is $1'\times4''$ (4.4 kpc $\times$ 0.29 kpc).
An HI profile is reproduced from Irwin \& Seaquist (1991).
The CO profile is rather round, while the HI profile has double-horns.
The CO width (${W_{20}^{\rm CO}=560}\pm$10 km/s)
is  broader than that of HI
(${W_{20}^{\rm HI}=510}\pm10$ \kms). The corresponding $\Delta D$
is 17\%.

NGC 3627 and 3628:
These two galaxies belong to the Leo Triplet and have been observed
in CO by Young et al. (1983). The authors also give a comparison
of the total line profiles of CO and HI for these two galaxies,
which are reproduced here. The good agreement between
 the profiles for both galaxies can clearly be seen. For
NGC 3627 we get
${W_{20}^{\rm CO}=380}\pm$20 km/s and ${W_{20}^{\rm HI}=385}\pm$20 km/s.
The linewidths of NGC 3628 are ${W_{20}^{\rm CO}=470}\pm$20 km/s and
{${W_{20}^{\rm HI}=480}\pm$20 km/s. This corresponds to
differences in distances
of -2\% for NGC 3627 and 3\% for NGC 3628 respectively.

NGC 4565:
This is  a ``standard'' edge-on galaxy of Sb type,
without any sign of interaction and warping.
Again, we find an almost identical line profiles both for CO and HI,
 even including the lobsidedness with respect to the systemic velocity.
Both the HI and  CO profiles are associated with
 a high-velocity wing at its 10--20\% level,
 with the CO wing being slightly stronger
 (${W_{20}^{\rm CO}=570}\pm1$km/s,
${W_{20}^{\rm HI}=530}\pm10$km/s). $\Delta D$ for this galaxy
is 13\%.

NGC 4594:
This edge-on galaxy has been observed by Bajaja et al. (1991).
It's morphological type is most likely to be Sa and the total
linwidths of these galaxy are extremely wide.
 Superimposed on the CO spectrum is the result
of an HI observation from Bajaja et al. (1984).
The total linewidths of CO and HI agree pretty well
(${W_{20}^{\rm CO}=750}\pm20$km/s,
${W_{20}^{\rm HI}=790}\pm20$km/s). This leads
to a difference in resulting distance of -8\%.

NGC 4631:
This is a  peculiar and interacting edge-on galaxy of Sc type,
 and its morphology is rather amorphous  (Sandage 1961).
Both the HI and CO profiles have a similar
double-horn + central peak structure.
 The HI systemic velocity is blue-shifted by
 about 20 km/s from CO, which may be due to a
tidal disturbance on the outer HI gas (Weliachew et al. 1978).
The HI line width (${W_{20}^{\rm HI}=330}\pm10$km/s) is
 broader than CO (${W_{20}^{\rm CO}300}\pm$10 \kms),
 which may be also due to tidal disturbance on the HI envelope.
 $\Delta  D$ for this Galaxy is -15\%.

NGC 4736:
The linewidths of CO (${W_{20}=250}\pm20$ km/s) and HI
(${W_{20}=235}\pm20$)
agree very well for this Sab galaxy
which has been observed in CO by Garman \& Young (1986) and in HI
by Rots (1980). The corresponding difference in distances is 10\%.

NGC 5194 (M51):
The CO profile (${W_{20}=180}\pm20$ km/s)
from Scoville and Young (1983)
and the corresponding HI profile
(${W_{20}=190}\pm10$ km/s) from Rots (1983)
show a good agreement despite the peculiarities of this galaxy like for
example the close companion NGC 5195. CO and HI distances differ by
$\Delta  D =-9\%$.

NGC 5457 (M 101):
This giant  Sc galaxy is one of the largest and nearest
face-on late-type spirals. CO (${W_{20}=150}\pm20$ km/s and
HI (${W_{20}=190}\pm20$ km/s) profiles differ
significantly in their width for
this galaxy. The CO profile is taken from Solomon et al. (1983),
the HI profile comes from Rots (1983). The CO and HI distances therefore
differ by $\Delta D = -33\%$ for this galaxy.

NGC 5907:
This edge-on Sc galaxy has been observed with the  Nobeyama 45m-telescope
in December 1992. The total line profiles of CO and HI (Rots 1983)
both show a
double horned structure  and coincide
almost perfect (${W_{20}^{\rm CO}=480}\pm20$ km/s,
${W_{20}^{\rm HI}=490}\pm20$ km/s). This difference
 between the linewiths corresponds
to a $\Delta D$
 of only -3\%.

NGC 6643:
The CO observation of this SAc galaxy was taken from Sanders \&
Mirabel (1985), the HI profile
was taken from Staveley-Smith (1988). The areement between
CO (${W_{20}=330}\pm20$ km/s) and HI (${W_{20}=340}\pm10$ km/s)
is pretty good, $\Delta  D$ is -5\%.

NGC 6946 and IC 342:
These rather face-on
Scd galaxies have been mapped in CO by Young \& Scoville (1982),
the HI data was taken from Rots (1979).
 For NGC 6946 the agreement is pretty good with a slightly broader
profile in HI  (${W_{20}^{\rm CO}=255}\pm 10$ km/s,
${W_{20}^{\rm HI}=260}\pm 20$ km/s).
IC 342 also shows a  broader HI profile but with a bigger difference
between the linewidths
(${W_{20}^{\rm CO}=150}\pm 20$ km/s,
${W_{20}^{\rm HI}=190}\pm 10$ km/s).
This difference between the linewidths
corresponds to a $\Delta D$ of
-33\%
for IC 342 and 3\% for NGC 6946 respectively.

NGC 7479:
A Sb galaxy with a broader HI profile (${W_{20}=370}\pm10$ km/s)
compared to CO (${W_{20}=310}\pm20$ km/s).
The CO profile was taken from Young \& Scoville (1983)
and the HI profile was taken from Staveley-Smith et al. (1988).
For this galaxy $\Delta D$ is  23\%.

UGC 8058 (Markarian 231):
A comparison of CO and HI profiles for this galaxy with a very
high infrared luminosity of more than $10^{12}$ \lsun \  has
been published by Sanders et al. (1987).
HI is observed in Absorption but the linewidths
(${W_{20}=230}\pm20 $km/s for HI,
${W_{20}=240}\pm20$ km/s for CO)
are very similar.
No inclination is given for this galaxy
but the good agreement between CO and HI would lead to no notable
difference in distances.

I ZW 1:
This is a quasar host galaxy and the CO data has been taken
from Barvainis and Alloin (1989). The HI profile is from
Condon et al. ( 1985) and it coincides very well with
the corresponding CO profile. The linewidths are
 ${W_{20}^{\rm CO}=405}\pm 20$ km/s and
${W_{20}^{\rm HI}=405}\pm 20$ km/s.
It is the most distant galaxy of our sample with a systemic velocity
of 18313 km/s. Since there is no inclination measurement for this
galaxy we can not use the \tfr to determine the distance. Hence, there
 would also be no notable difference between
CO and HI distances.\v
\no
\cen{-- Figure 1 --}
\v
\no

\no
{\bf 3. Results}\v

Table 1 shows a compilation of the measured linewidths for our
sample of galaxies. Column one gives the galaxy name, column two
the linewidths of CO and HI at the 20\% level of peak intensity
and column four and five at the 50\% level respectively.

Figure 2 shows a plot of the HI-linewidths versus the
CO-linewidths at the 20\% and
at the 50\% level.
The correlation is better
at the 20\% level with ${W^{\rm CO}/W^{\rm HI}=0.96}\pm$0.10, whereas for the
50\% level we found ${W^{\rm CO}/W^{\rm HI}=1.03}\pm$0.21.
\v
\no
\cen{-- Figure 2 --}
\v

Finally we give the distances to the galaxies of our sample
using the linewidth measured in HI and secondly the distances
resulting from the CO Tully-Fisher relation.
 We thereby apply the most recent
calibration of the B-band \tfr given by Pierce and Tully (1992):
$${M_{B}=-7.48\ ({\rm log} W^{i}-2.50)-19.55+\delta m_{B}}$$
Log $W^{i}$ are the logarithms of the linewidths corrected for
internal turbulence (Tully and Fouqu\'e 1985) and inclination.
$\delta m_{B}$ is a correction term for cluster galaxies which
has not to be applied in our case. To calculate the distances
we used the apparent blue magnitudes
which are listed in table 1 and the absolut magnitudes determined by
the calibration of the \tfr and our measured linewidths.

Table 2 shows a compilation of the result.
The first column gives
 the galaxy name, the second the distance derived from the HI
Tully-Fisher relation,
 the third the distance from the CO \tfr and column four shows
  the deviation of the CO distances from that of HI.
\v
\no
Table 2
\v

In figure 3
the CO distances are plotted against the HI distances.
The scatter for the distances
 is slightly broader than for
the linewidths,
we get a value of
${D_{\rm CO}/D_{\rm HI}= 0.94}\pm0.16$.

 Figure 3 also shows the correlation for
the galaxies with distances of up to 15 Mpc. In this
plot the very good correlation is obvious.
\v
\no
\cen{-- Figure 3 --}

\v

\no{\bf 4. Discussion}

\v
\no{\it 4.1. Deviation between CO and HI profiles for some galaxies}\v

Although most of the sample galaxies showed good
correlation between CO and HI, some cases were found in
which they do not coincide sufficiently.
In the following we make some remarks about
possible reasons for larger deviations from the
overall correlation between CO and HI.

\no
Tidal interactions:

It is interesting that most of the galaxies where the differences
in resulting distances from using CO instead of HI exceed 10\%
are suspected to be interacting with a companion
(Condon \& Condon 1982). This suggests that interactions play
a role in disturbing the reliability of the \tfr\ as a distance indicator.
Since CO is more concentrated to the center than HI it is not as
sensitive to tidal interactions as HI. Therefore the inrinsic scatter
of the CO \tfr might even be better than for HI and it could even be
a usefull tool not only for distant but also for nearby galaxies.

However, from our data  we can not decide whether the linewidths of
CO or HI are better for determining the rotational characteristics
of a galaxy, since the distances so far obtained for these galaxies
are too crude to calibrate the \tfr\.
To look at the intrinsic scatter of the CO \tfr
further observations of clusters of galaxies are necessary.

\no
Galactic contamination:

In the case of IC 342 the rather large difference between
the CO and HI distances of -33\%  might be due to galactic
contamination, since the
systemic velocity of this galaxy
is close to zero.
Another explanation also might be the interaction with a companion
(Rots 1979).\v

\v\no{\it 4.2. Normal vs peculiar galaxies}

Among the 32 galaxies in the sample, there are two with large
systemic velocities (NGC 2623 and Izw 1), both of which
are galaxies with unusual activity.
NGC 3034 (M82) is a peculiar starburst galaxy, and NGC 1068
shows also starburst activity.
It is rather surprising that the total line widths (including
absorption for HI) for HI and CO in these peculiar
cases show a good coincidence.
This fact provides an interesting case about
studying the rotation and interstellar properties of such peculiar active
galaxies, besides the interest in \tfr.
A thourough investigation for the reason of the
agreement, paraticularly for the former two cases,
is needed by obtaining higher-resolution imaging data
in CO and HI.

Moreover, our result gives a clear answer to the current
guess that CO must be concentrated in the
central region of galaxies, so that its line profiles would
be significantly different from HI, and would be not adequate for
such a purpose to measure rotation velocities.
However, this is not the case as Fig. 1 indicates.
The reason for the HI-CO coincidence has been
investigated by Sofue (1992), who examined the
rotation characteristics and molecular hydrogen and HI distributions
in the well-resolved edge-on galaxies NGC 891 and NGC 4565.

\v\no{\it 4.3. On the Use of CO \tfr}

The \tfr\ makes use of the properties of normal galaxies.
Except for some galaxies as raised above, most of the
sample galaxies are normal spirals.
The accuracy of velocity width measurements is usually
$\pm 10$ \kms in CO-line observations, and is comparable
with that of HI.
Moreover, the CO data which we demonstrated here are
taken (or reconstructed using data) from the literature
except for several galaxies from the 45-m telescope,
and have not been obtained properly for the purpose of
distance measurements.
Nevertheless, we obtained a good correlation.
Hence, by a systematic measruement program
of CO-linewidth using adequate telescopes
equiped with the best receivers, CO width
data of accuracy $\pm 5$ \kms would be routinely
obtained by integration time of a few hours
for galaxies below $cz \sim 10,000$ \kms.
For \tfr\ of more distant cluster galaxies,
to which HI cannot reach, we certainly need the
largest facilities such as the Nobeyama 45-m telescope
and the IRAM 30-m telescope.

It would be worthwhile to demonstrate that CO \tfr
measurement is quite possible for distant cluster galaxies
using the 45-m telescope:
Normal Sb and Sc galaxies of NGC 891 size
contain molecular hydrogen in
mass of the order of $10^{9-10}\Msun$, which corresponds to
CO ($J=1-0)$ line intensity of
$I_{\rm CO} \simeq 4 (D/100{\rm Mpc})^{-2}$ K \kms
with the 45-m telescope (beam width 15$''$)
for a normal CO-to-H$_2$ conversion
factor (Bloemen et al. 1986).
For a typical line width of about 400 \kms, we may, thus, expect
a main-beam brightness temperature of about
$\Tmb \simeq 10-100 (D/100{\rm Mpc})^{-2}$ mK.
This implies that measurements with rms noise of a few mK
are required for a velocity resolution of 10 \kms,
which is quite possible by an integration
time of a few hours in the present status of telescope
under a good weather condition.
If the system noise temperature decreases by a factor of ten,
due to improvements of receivers as well as due to
larger redshift of galaxies, we may be able to obtaine
CO line-width data rather ``routinely'', as is usually done
in HI for nearby galaxies.
In this respect, we have been indeed starting a long-term project of
CO-line observations for the \tfr\ using the 45-m telescope at Nobeyama.

\no{\bf 5. Conclusion}\v

We compared the total line profiles of CO and HI for
a sample of 32 galaxies.
Confirming the work of
Dickey \& Kazes (1992) and Sofue (1992)
we also found a correlation between
CO and HI.
However, for the galaxies of our
sample which are supposed to be interacting
 we obtained larger differences between CO and HI than
they found in their sample.
This might indicate the presumably better reliability
of the CO linewidth compared to HI for interacting galaxies.
In order to confirm this further studies of cluster galaxies at the same
distance are necessary  to look at the intrinsic
scatter of the CO-\tfr.

Since the mean value of ${D_{\rm HI}/D_{\rm CO}}$ is close to one and the
scatter relatively low, our study shows that CO is a promising
tool, even if the intrinsic scatter proves not to be
significantly better than for the HI \tfr.

\v
\no
{\it Acknowledgements.} F. Sch\"oniger receives support
from the Japanese-German Center Berlin and also wants
to express his gratitude to the staff of the University
of Tokyo
and Nobeyama Radio Observatory for their hospitality.
\vfill
\eject

\sect{\bf References} \v

\r Aaronson M., Bothun G., Mould J., Shommer
   R. A., Cornell, M. E., 1986, ApJ 302, 536

\r Bajaja E., vanden Burg G., Faber S. M., et al., 1984,  A\&A 141, 309

\r Bajaja E., Krause M., Dettmar R.-J., Wielebinski R., 1991, A\&A 241, 411

\r Barvainis R., Alloin D., Antonucci R., 1989, ApJ 337, L69

\r Bottinelli L., Fouqu{\'e} P., Gouguenheim L.,
     Paturel G., 1990, A\&A 82, 391 

\r Bloemen, J. B. G. M., Strong, A. W., Blitz, L., Cohen, R. S.,
   Dame, T. M., Grabelsky, D. A., Hermsen, W., Lebrun, F.,
   Mayer-Hasselwander, H. A., and Thaddeus, P. 1985, \aa, 154, 25.

\r Condon J. J., Condon M. A., Gisler G., Puschell J. J., 1982, ApJ 252, 102

\r Condon J. J., Hutchings J. B.,  Gower A. C., 1985, AJ 90, 1642

\r Crutcher R. M., Rogstad D. H.,  Chu K., 1978,
    ApJ 225, 784

\r Davies R. D.,  Staveley-Smith L., Murray J. D., 1989, MNRAS 233, 174

\r de Vaucouleurs G., de Vaucouleurs A., Corwin  H. G. Jr., et al.,
   1991, in
   {\it  Third Reference Catalogue of Bright Galaxies}
   (New York: Springer Verlag)

\r Dickey J., Kazes I., 1992, ApJ 393, 530

\r Duric N., Seaquist E. R., Crane P., Bignell R. C.,  Davis L. E., 1983,
    ApJ 273, L11

\r Fukugita M., Okamura S., Tarusawa K., et al., 1991, ApJ 376, 8

\r Fouqu{\'e} P., Bottinelli L., Gouguenheim L., 1990, ApJ 349, 1

\r Garman L. E., Young J. S., 1986, A\&A 154, 8

\r Huchtmeier W., K., Richter O.-G., 1989, in {\it A General Catalog
  of HI Observations of Galaxies, Springer-Verlag, Heidelberg,  Table 1}

\r Irwin J., Seaquist E.  R., 1991, ApJ 371, 111

\r Knapp G. R., Shane W. W., 1984, A\&A 141, 309

\r Koper E., Dame T. M., Israel F. P., Thaddeus P., 1991, ApJ 383, L11

\r Kraan-Korteweg R. C., Cameron L. M.,  Tammann G. A., 1988, ApJ 331, 610

\r Mirabel I. F., Sanders D. B., 1988, ApJ 335, 104

\r Pierce M. J., Tully R. B., 1988, ApJ 330, 579

\r Pierce M. J., Tully R. B., 1992, ApJ 387, 47

\r Rots A., 1979, A\&A 80, 250

\r Rots, A., 1980, A\&AS, {\bf 41}, 189

\r Sandage A. R., 1961, {\it The Hubble Atlas of Galaxies} (Carnegie
  Institution, Washington),  25

\r Sandqvist Aa., Elfhag T., J\"ors\"ater S., 1988, A\&A 201,
     223

\r Sanders D. B., Mirabel I. F., 1985, ApJ 298, L31

\r Sanders D. B., Young J. S., Scoville N. Z., et al., 1987, ApJ 312, L5

\r Scoville N. Z., Young J. S., 1983, ApJ 265, 148

\r Scoville N. Z., Young J. S., Lucy L. B., 1983, ApJ 270, 443

\r Shostak G. S., 1978, A\&A  68, 321

\r Sofue Y., 1992, PASJ 44, L231

\r Sofue Y., Handa T., Nakai N.,   1989, PASJ 41, 937

\r Sofue Y., Taniguchi Y., Wakamatsu K., Nakai N., 1993, PASJ in press

\r Sofue Y., Irwin J., 1992, PASJ 42, 353

\r Sofue Y., Nakai N., 1993,  PASJ, in press

\r Solomon P. M., Barret J., Sanders D. B.,  de Zafra R., 1983,
      ApJ 266, L 103

\r Staveley-Smith L., Davies R. D., 1988, MNRAS 231,  833

\r Tift H. W., Cocke W. G., 1988, ApJS 67, 1

\r Tully B., Fisher J. R., 1977, A\&A 64, 661

\r Tully B., Fouqu{\'e} P., 1985, ApJS 58, 67

\r V{\'e}ron-Cetty M.-P.,  V{\'e}ron P., 1985, A\&A 145, 425

\r Weliachew L., Sancisi R.,  Guelin M., 1978,  A\&A 65, 37

\r Young J. S., Tacconi L. J., Scoville N. Z., 1983, ApJ 269, 136

\r Young J. S., Scoville, N. Z., 1982,  ApJ 258, 467

\r Young J. S., Scoville, N. Z., 1984,  ApJ 287, 153

\r Young J S., Gallagher J. S., Hunter D. H., 1984, ApJ 276, 476

\r Young J. S., Schloerb F. P., Kenney J. D.,
   Lord S. D., 1986, ApJ 304, 443

\vfill
\eject
\no
$$\vbox {\halign {#\hfil && \quad \hfil #\hfil \cr
\multispan7 {\bf Table 1.} Galaxy properties and measured linewidths \hfil &
\cr
\noalign{\smallskip} \cr
\noalign{\hrule} \cr
\noalign{\smallskip}\cr
Galaxy & Inclination [\deg] & $B_{0}^{\rm T}$ [mag]
 & $W_{20}^{\rm CO}$ [km/s] & $ W_{20}^{\rm HI}$ [km/s] &
$ W_{50}^{\rm CO}$ [km/s] & $ W_{50}^{\rm HI}$  [km/s] \cr
\noalign{\smallskip}\cr
\noalign{\hrule}\cr
NGC 224  & 75 & 3.34  & 535 & 540 & 510 & 510 \cr
NGC 520  & 68 & 11.77 & 360 & 420 & 280 & 150 \cr
NGC 660  & 66 & 11.44 & 370 & 330 & 305 & 335 \cr
NGC  891 & 85 &  9.5  & 490 & 490 & 460 & 460 \cr
NGC 992  & 47 & 13.86 & 385 & 360 & 300 & 240 \cr
NGC 1068 & 40 & 9.46  & 350 & 345 & 290 & 260 \cr
NGC 1365 & 61 &  9.90 & 370 & 390 & 270 & 370 \cr
NGC 1569 & 63 &  9.45 & 85  & 95  & 70  & 85  \cr
NGC 1808 & 57 & 10.47 & 340 & 325 & 315 & 250 \cr
NGC 2146 & 34 & 10.52 & 390 & 460 & 300 & 260 \cr
NGC 2276 & 27 & 11.44 & 140 & 200 &  95 & 95  \cr
NGC 2339 & 43 & 11.54 & 330 & 340 & 310 & 330 \cr
NGC 2623 & 64 & 13.19 & 450 & 400 & 400 & 360 \cr
NGC 3034 & 82 & 8.83  & 320 & 300 & 250 & 200 \cr
NGC 3079 & 85 & 10.45 & 560 & 510 & 530 & 450 \cr
NGC 3627 & 62 & 9.13  & 380 & 385 & 310 & 310 \cr
NGC 3628 & 82 &  9.32 & 470 & 480 & 330 & 350 \cr
NGC 4565 & 90 & 9.11  & 570 & 530 & 500 & 500 \cr
NGC 4594 & 85 & 8.39  & 750 & 790 & 750 & 770 \cr
NGC 4631 & 85 & 8.61  & 305 & 330 & 260 & 300 \cr
NGC 4736 & 35 & 8.72  & 250 & 235 & 230 & 220 \cr
NGC 5194 & 20 & 8.68  & 180 & 190 & 160 & 140 \cr
NGC 5457 & 30 & 8.19  & 150 & 190 & 140 & 170 \cr
NGC 5907 & 90 & 9.74  & 480 & 490 & 455 & 470 \cr
NGC 6643 & 60 & 11.19 & 330 & 340 & 230 & 320 \cr
NGC 6946 & 30 & 7.84  & 255 & 260 & 205 & 230 \cr
NGC 7469 & 45 & 12.75 & 330 & 370 & 245 & 230 \cr
NGC 7479 & 39 & 11.33 & 310 & 370 & 250 & 350 \cr
NGC 7674 & 25 & 13.55 & 190 & 215 & 160 & 160 \cr
IC 342   & 25 & 5.58  & 150 & 190 & 125 & 170 \cr
Mrk 231  &    & 13.97 & 240 & 230 & 180 & 160 \cr
I Zw 1   &    & 13.99 & 405 & 405 & 400 & 400 \cr
\noalign{\smallskip}\cr
\noalign{\hrule}\cr
}}$$
\vfill
\eject
\no
$$\vbox {\rm \halign{#\hfil && \quad \hfil #\hfil \cr
\multispan4 {\bf Table 2.} Resulting Distances \hfil & \cr
\noalign{\smallskip} \cr
\noalign{\hrule}\cr
\noalign{\smallskip}\cr
Galaxy &
$ D_{\rm CO}$ [Mpc] & $ D_{\rm HI}$ [Mpc]
   & ${ \Delta D=(D_{\rm CO}-D_{\rm HI})/{D_{\rm HI}}}$ [\%]\cr
\noalign{\smallskip}\cr
\noalign{\hrule}\cr
NGC  891 & 11.3 & 11.3 & 0 \cr
I Zw 1   &       &      & 0 \cr
Mrk 231  &      &      & 0 \cr
NGC 224  & 0.77    & 0.78  &  $-1$ \cr
NGC 1068 & 12.6 & 12.3 & 2 \cr
NGC 3627 &  7.3 &  7.4 & $-2$ \cr
NGC 3628 &  9.4 &  9.7 & $-3$ \cr
NGC 6946 &  5.4 &  5.5 & $-3$ \cr
NGC 5907 & 11.6 & 12.0 & $-3$ \cr
NGC 2339 & 27.0 & 28.3 & $-5$ \cr
NGC 6643 & 15.4 & 16.2 & $-5$ \cr
NGC 1808 & 12.3 & 11.7 & 5 \cr
NGC 4594 & 12.9 & 14.0 & $-8$ \cr
NGC 5194 & 8.3  &  9.1 & 9 \cr
NGC 1365 & 8.0 & 8.7 & $-9$ \cr
NGC 4736 & 6.2   & 5.6  & 10 \cr
NGC 992  & 90.2 & 80.8 & 11 \cr
NGC 3034 & 3.9 & 3.5 &  11 \cr
NGC 4565 & 11.5 & 10.2 & 13 \cr
NGC 4631 & 3.2   & 3.7 & 15  \cr
NGC 7469 & 44.4 & 53.6 & $-17$ \cr
NGC 3079 & 20.9 & 17.9  & 17 \cr
NGC 7674 & 60.3 & 73.9 & $-18$ \cr
NGC 660  & 19.2  & 15.8 & 22 \cr
NGC 2623 & 61.1 & 50.0 & 22 \cr
NGC 520 & 20.8 & 26.9 &  $-23$ \cr
NGC 2146 & 30.5 & 39.7  & $-23$ \cr
NGC 7479 & 25.2 & 33.7  & 25 \cr
NGC 1569 & 0.4 & 0.5  & $-25$ \cr
NGC 5457 & 2.6  &  3.9  & 33 \cr
IC 342   & 1.0 & 1.5   & $-33$ \cr
NGC 2276 & 12.0 & 22.1 & $-36$ \cr
\noalign{\smallskip}\cr
\noalign{\hrule}\cr
}}$$
\no

\vfill
\eject
\v
\no
Figure Captions:
\v
\no
{\bf Fig. 1.} The total line profiles of CO (solid lines)
 and HI (dashed lines) for our sample of galaxies
 \vskip 10mm

\no
 {\bf Fig. 2a and b.} CO linewidths of CO plotted against the corresponding
 HI linewidths (a) at the 20\% level  and (b) at the 50\% level.
 \vskip 10mm

\no
 {\bf Fig. 3.} The CO distances plotted against the corresponding
 HI distances (a) for all galaxies  and (b) for the galaxies
 with distances
 up to 15 Mpc.

\endpage
\bye
\end